\documentclass[cits,a4paper]{PoS}

\newcommand{\iso}[2]{\hbox{${}^{#1}{\rm #2}$}}

\title{The impact of the $^{18}$F($\alpha$,$p$)$^{21}$Ne reaction on fluorine production in 
AGB stars}

\ShortTitle{The $^{18}$F($\alpha$,$p$) reaction and AGB stars}

\author{\speaker{Amanda I. Karakas}$^{a}$, H. Y. Lee$^{b,d}$, Maria Lugaro$^{c}$,
        J. G\"{o}rres$^{d}$ and M. Wiescher$^{d}$\\
        \llap{$a$} Research School of Astronomy \& Astrophysics, Mt Stromlo Observatory,
Weston Creek ACT 2611, Australia\\ 
        \llap{$b$} Physics Division, Argonne National Laboratory, Argonne, IL 60439-4843\\
        \llap{$c$} Sterrenkundig Instituut, University of Utrecht, Postbus 80000, 3508 TA Utrecht, The Netherlands\\
        \llap{$d$} Department of Physics and Joint Institute for Nuclear Astrophysics, 
	University of Notre Dame, Notre Dame, IN 46556\\
E-mail: \email{akarakas@mso.anu.edu.au}, \email{hylee@phy.anl.gov}, \email{M.Lugaro@phys.uu.nl}
	\email{jgoerres@nd.edu,mwiesche@nd.edu}}

\abstract{The recent experimental evaluation of the 
$^{18}$F($\alpha$,$p$)$^{21}$Ne reaction rate, when considering its 
associated uncertainties, presented significant differences compared 
to the theoretical Hauser-Feshbach rate.  This was most 
apparent at the low temperatures relevant for He-shell burning in 
asymptotic giant branch (AGB) stars. Investigations into the effect on 
AGB nucleosynthesis revealed that the upper limit resulted in an enhanced
production of $^{19}$F and $^{21}$Ne in carbon-rich AGB models, but
the recommended and lower limits presented no differences from using
the theoretical rate. This was the case for models spanning a range
in metallicity from solar to [Fe/H] $\sim -2.3$. 
The results of this study are relevant for observations of F and 
C-enriched AGB stars in the Galaxy, and to the Ne composition of mainstream 
silicon carbide grains, that supposedly formed in the outflows of cool, 
carbon-rich giant stars. We discuss the mechanism that produces the 
extra F and summarize our main findings.} 

\FullConference{10th Symposium on Nuclei in the Cosmos\\
		 July 27 - August 1 2008\\
		 Mackinac Island, Michigan, USA}

\begin{document}

\bibliographystyle{unsrt}

\section{Introduction}

Until 2006 the only rate for the $^{18}$F($\alpha$,p)$^{21}$Ne reaction 
was the theoretical estimate available in the Brussels nuclear
reaction-rate library \cite{aikawa05}.  The first experiment 
aimed at determining the $^{18}$F($\alpha$,p)$^{21}$Ne rate 
over a large range of stellar temperatures was carried out by
Lee \cite{LeeThesis}.
This experimental evaluation, when considering its associated 
uncertainties, presented significant differences compared to the 
theoretical rate, especially at the low temperatures relevant for
He-shell burning in asymptotic giant branch (AGB) stars 
($T \approx 0.3$~GK). We investigate the effect of such 
differences on the nucleosynthesis occurring in AGB models of various
initial mass and composition.

The theoretical estimate of the $^{18}$F($\alpha$,p)$^{21}$Ne rate
was not present in previous studies e.g., \cite{karakas06a}, 
although we had included the species \iso{18}F because of its 
important role in the reaction chain 
\iso{14}N($\alpha,\gamma$)\iso{18}F($\beta^+\nu$)\iso{18}O,
leading to the production of \iso{18}O in the He shell. 
It was found that the inclusion of the $^{18}$F($\alpha$,p)
reaction resulted in an
increase in the production of the stable \iso{19}F; see \S\ref{results}
for more details on the production mechanism. 
This is of interest because AGB models do not synthesize enough 
$^{19}$F to match the [F/O] abundances observed in AGB stars 
\cite{jorissen92}. Also, the cosmic origin of fluorine is still 
uncertain, with massive stars \cite{woosley95} 
playing a significant role in producing fluorine alongside AGB stars.
However, AGB stars and their progeny (e.g., 
post-AGB stars, planetary nebulae) are still the only confirmed
site of fluorine production thus far \cite{federman05,
werner05}. Observations of an enhanced F abundance 
([F/Fe] = 2.90) in a carbon-enhanced metal-poor halo star 
\cite{schuler07} is further motivation to understand the 
details of F production in AGB stars \cite{lugaro08}.

In this proceedings, we summarize the results of calculations
published in Karakas et al. \cite{karakas08a}.

\section{The $^{18}$F($\alpha$,$p$)$^{21}$Ne reaction rate}

The measurement of the $^{18}$F($\alpha$,p)$^{21}$Ne reaction cross 
section is made difficult by the short half-life 
($T_{\rm 1/2} \sim$~109 min) of \iso{18}F. The time-reversed reaction of 
\iso{21}Ne($p,\alpha$)\iso{18}F was investigated at the Nuclear Science
Laboratory in the University of Notre Dame \cite{LeeThesis}. 
The cross section was measured in the energy range of 2.3 MeV 
to 4.0 MeV using the activation method. The lower limit of the 
cross-section measurement is mainly determined by the statistical 
uncertainty of the activation data, while the upper limit is based 
on the uncertainty associated 
with the $^{18}$O induced background.  We refer the reader 
to \cite{karakas08a} for further details.

\section{Results}  \label{results}

We computed the stellar structure first using 
the Mt Stromlo Stellar Structure code, and then 
performed post-processing on that structure to obtain abundances 
for 77 species, most of which are not included in the small 
stellar-structure network. See \cite{karakas08a} and references
therein for more details.
This technique is valid for studying reactions not directly related 
to the main energy generation. This is certainly the case for 
studying the effect of the $^{18}$F($\alpha,p$)$^{21}$Ne 
reaction on AGB nucleosynthesis. 
We included models in our study with masses between 1.9 to 
5$M_{\odot}$, with initial metallicities from solar ($Z=0.02$) 
to $Z=0.0001$ ([Fe/H] $-2.3$), computed previously in 
\cite{karakas07b}.
A partial mixing zone (PMZ) is required to produce a $^{13}$C 
{\it pocket} and free neutrons in the He-intershell 
via the $^{13}$C($\alpha,n$)$^{16}$O
reaction. Neutrons are necessary for the chain \iso{14}N($n,p$)\iso{14}C,
where free protons are then used by \iso{18}O($p,\alpha$)\iso{15}N.
We include a PMZ of 0.002$M_{\odot}$ for all  lower mass cases.
The protons in the PMZ are captured by the abundant \iso{12}C to form
a \iso{13}C pocket  that is $\approx 10-15$\% of the mass of the He-intershell. 
We note that the 
extent in mass and the proton profile of the partial mixing zone are 
very uncertain parameters (see discussions in \cite{herwig05,lugaro04}).

\begin{table}[t]
\begin{center}
\caption{Results from the AGB models. For each mass and $Z$ value, 
we show the mass of the partial mixing zone used 
in the computation, the yield ($y$) of $^{19}$F, and the multiplication 
factor ($X$) needed to obtain the upper limit $^{19}$F yield from 
the recommended-rate yield. All yields are in solar masses, and the
multiplication factors are dimensionless quantities.
The same information is also presented for $^{21}$Ne for each model.
\label{tab:yields}}
\vspace{1mm}
\begin{tabular}{@{}ccccccc@{}} 
\hline\hline
 Mass &  $Z$ & PMZ & $y$($^{19}$F$_{\rm rec}$) & $X$($^{19}$F) &
 $y$($^{21}$Ne$_{\rm rec}$) & $X$($^{21}$Ne) \\
\hline
 3.0  &  0.02  & 0.002  & 5.84($-6$) & 1.526 & 1.25($-6$) & 4.423 \\
 3.0  &  0.012 & 0.002  & 5.66($-6$) & 1.736 & 1.39($-6$) & 5.330 \\
 1.9  &  0.008 & 0.002  & 9.35($-7$) & 1.178 & 1.60($-7$) & 2.340 \\
 3.0  &  0.008 & 0.002  & 1.71($-5$) & 2.407 & 4.52($-6$) & 9.609 \\
 2.5  &  0.004 & 0.002  & 1.33($-5$) & 2.061 & 2.81($-6$) & 8.364 \\
 5.0  &  0.004 & 0      & 1.45($-7$) & 4.582 & $-$2.58($-6$) & $-$1.965 \\
 2.0  & 0.0001 & 0.002  & 1.67($-5$) & 1.975 & 3.23($-6$) & 8.551 \\   \hline
\end{tabular}

\end{center}
\end{table}

From Table~\ref{tab:yields} it is evident that employing the 
upper limit of the $^{18}$F($\alpha,p$)$^{21}$Ne reaction results in an
increase in the production of $^{19}$F and $^{21}$Ne, compared
to using the recommended rate. 
The change in the yield increases with decreasing metallicity, 
at a given mass, with the largest change found in the 
5$M_{\odot}$, $Z=0.004$  model.  Note that the 
amount of $^{19}$F produced in the intermediate-mass models
(at a given $Z$) is much less than the amount produced in 
the lower-mass 3$M_{\odot}$ model, by factors of 
$\sim$3--40. This is because $^{19}$F is destroyed by HBB 
in the 5$M_{\odot}$ models.  

The enhanced abundance of $^{19}$F as a result of using the
upper limit  may be explained by considering
the $^{18}$O($p, \alpha$)$^{15}$N($\alpha,\gamma)^{19}$F reaction
chain. Including the $^{18}$F($\alpha, p$)$^{21}$Ne reaction
reduces the abundance of $^{18}$O because it competes with
$^{18}$O production via the $^{18}$F($\beta^+\nu)^{18}$O decay.
However, the extra amount of protons from ($\alpha, p$) enhances
the $^{18}$O($p, \alpha$)$^{15}$N reaction rate, even though
$^{18}$O production has been deprived from the decay.
In other words, the sum $N_{^{18}{\rm O}}+N_{p}$
(where $N_{i}$ is the abundance by number of nucleus $i$) remains
constant, however, the product $N_{^{18}{\rm O}} N_{p}$, on
which the number of $^{18}$O+$p$ reactions depends, is maximized
when $N_{^{18}{\rm O}}$ is equal to $N_{p}$.
In \cite{karakas08a} we analyzed the effect of 
the extra protons on the $^{19}$F production in the He-shell.
It was found that
the overall \iso{19}F production increases as long as 
$N_{^{14}{\rm N}}/N_{{p}_0} > 1$, where 
$N_{{p}_0}$ is the original number density of protons
without the inclusion of the \iso{18}F($\alpha,p$)\iso{21}Ne
reaction, and this condition is well 
satisfied in the He-burning shell. During the network calculation a 
realistic $N_{^{14}{\rm N}}/N_{{p}_0} \approx 10^{10}$; 
this ratio is large enough to explain the enhanced fluorine 
production in the stellar models.

\section{Discussion and conclusions}

\begin{figure}
\begin{center}
\includegraphics[width=0.7\textwidth]{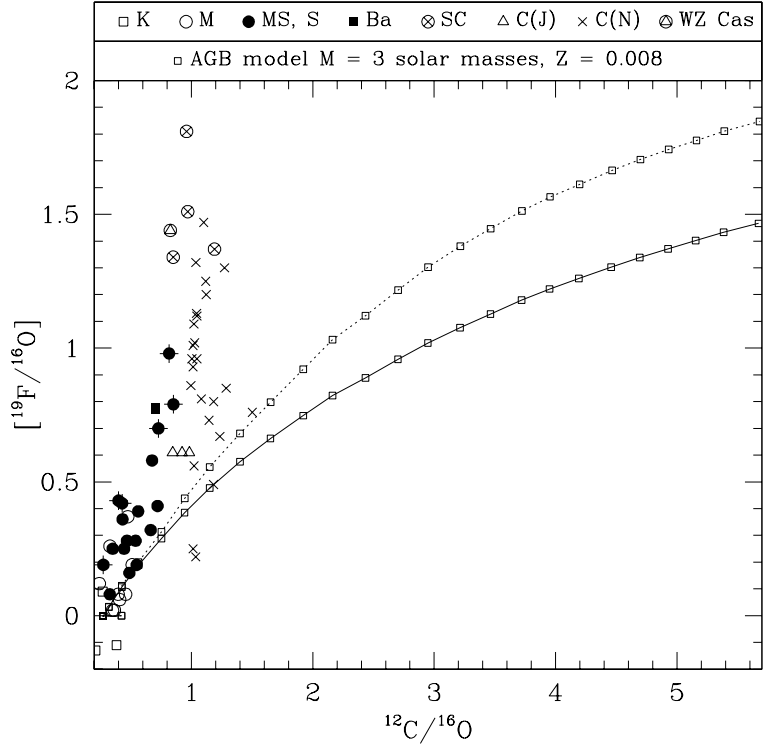}
\caption{Comparison of fluorine abundances observed by \cite{jorissen92}
and model predictions for the 3$M_{\odot}$, $Z=0.008$ model. 
The predictions are normalized such that the initial $^{19}$F abundance 
corresponds to the average F abundance observed in K and M stars. 
Each symbol on the prediction lines 
represents a TDU episode. Solid lines represent calculations performed 
using no $^{18}$F($\alpha,p$)$^{21}$Ne reaction, which are equivalent 
to using the current lower limit, recommended value and Brussels library 
rate. Dotted lines are calculations performed using the current upper 
limit of the rate.}
\end{center}
\label{fig1}
\end{figure}

From Fig.~\ref{fig1}, we see that the surface 
[$^{19}$F/$^{16}$O] ratios from the 3$M_{\odot}$, $Z=0.008$ 
model are a factor of $\sim 2.2$ times higher when 
employing the new upper limit of the 
$^{18}$F($\alpha,p$)$^{21}$Ne reaction. We chose to show
this model because it produces the largest F abundances, although
the metallicity of this model is probably at the lower end 
of the distribution of Galactic carbon stars. 
Table~\ref{tab:yields} and Fig.~\ref{fig1} shows that a 
match between the stellar models and the stars with the highest 
observed $^{19}$F abundances is possible, but only for 
very high C/O ratios of $\sim 4-5$ found in the 
3$M_{\odot}$, $Z = 0.008$ model.   These high C/O ratios
are likely not realistic, and the inclusion of carbon-rich, 
low-temperature opacities into the stellar models would 
cause the TP-AGB evolution to end before the model star
reached such C/O ratios. 
A solution to the mystery of the high F abundances at 
modest C/O ratios is still missing, but a re-evaluation
of the F and C abundances in the sample of AGB stars 
considered by \cite{jorissen92} may help, along with 
a detailed examination of model uncertainties.

The $^{18}$F($\alpha$,p)$^{21}$Ne reaction also affects 
the abundance of $^{21}$Ne in the He-shell of AGB stars.  
There is a long-standing puzzle concerning the isotopic 
composition of Ne measured in stellar silicon carbide (SiC) 
grains extracted from meteorites, which formed in the envelopes 
of carbon-rich AGB stars e.g., \cite{lewis90,zinner06}. Models 
computed with the upper limit of the 
$^{18}$F($\alpha, p$)$^{21}$Ne reaction rate show an 
increase in the $^{21}$Ne abundance, and hence in the 
$^{21}$Ne/$^{22}$Ne ratio in the intershell of up to a 
factor of 6; see \cite{karakas08a} for further details.  
We conclude that the \iso{18}F($\alpha,p$)\iso{21}Ne 
reaction rate being closer to its upper limit may be a 
promising explanation for the $^{21}$Ne/$^{22}$Ne ratios 
in SiC grains. The modeling uncertainties related to convection
and mass loss do not affect the intershell compositions 
and thus do not apply to the discussion 
of the Ne composition of stellar SiC grains. For this
reason, the results for Ne are also a more reliable hint
that the $^{18}$F($\alpha$,p)$^{21}$N reaction is 
indeed closer to its upper limit than the comparison 
of F in AGB stars.  However, more experimental data 
for this reaction at temperatures below 0.4~GK are 
required to help verify this result.  

\bibliography{apj-jour,library}


\end{document}